\documentclass{v16nufact}
\newcommand{\ii}{i}
\renewcommand{\Im}{{\rm Im}}
\renewcommand{\Re}{{\rm Re}}
\usepackage{upgreek}
\def\be{\begin{equation}}
\def\ee{\end{equation}}
\def\bea{\begin{eqnarray}}
\def\eea{\end{eqnarray}}

\newcommand{\Photo}{}

\begin{document}
\title{On the relation between the CP phases in the PMNS matrix, CP-violation with sterile neutrinos and leptogenesis}

\author{Marco Drewes$^1$, Bj\"orn~Garbrecht$^1$, Dario Gueter$^{1,2,3}$ and Juraj Klari\'{c}$^1$
}

\address{\footnotesize{$^1$Physik Department T70, Technische Universit\"at M\"unchen,}\\
\footnotesize{ James Franck Stra\ss e 1, D-85748 Garching, Germany}\\
\footnotesize{$^2$ Max-Planck-Institut f\"ur Physik (Werner-Heisenberg-Institut),}\\
\footnotesize{F\"ohringer Ring 6,
80805 M\"unchen, Germany}\\
\footnotesize{$^3$ Excellence Cluster Universe, Boltzmannstra{\ss}e 2,}\\
\footnotesize{Technische Universit\"at M\"unchen, 85748 Garching, Germany}
}

\maketitle
\begin{abstract}
We discuss the connection between observable CP violation in the lepton sector, the properties of heavy neutrinos and the baryon asymmetry of the universe in the minimal seesaw model. A measurement of the Dirac phase $\delta$ would allow to make testable predictions for the couplings of the heavy neutrinos to individual Standard Model lepton flavours. If any heavy neutral leptons are experimentally discovered in the future, this provides a powerful test for the mechanism of neutrino mass generation and baryogenesis.
\end{abstract} 

\section{Introduction} 
There is compelling evidence that the observable universe does not contain any significant amounts of antimatter at present time, and that the (baryonic) matter is the remnant of a tiny matter-antimatter asymmetry in the primordial plasma that filled the early universe 
that survived
after mutual annihilation of all particles and antiparticles, see e.g.\ Ref.~\cite{Canetti:2012zc} for a review.
This \emph{baryon asymmetry of the universe} (BAU) is commonly expressed in terms of the net baryon-to-photon ratio $\eta_B=n_B/n_\gamma\simeq 6\times 10^{-10}$. 
In most inflationary models, the radiation dominated epoch of cosmic history starts with $\eta_B=0$. 
The generation of a non-zero $\eta_B$ requires to fulfil the three well-known Sakharov conditions \cite{Sakharov:1967dj} $i)$ violation of baryon number $B$, $ii)$ violation of $C$ and $CP$ and $iii)$ a deviation from thermal equilibrium. 
While all three conditions are in principle met in the Standard Model (SM) of particle physics, the violation of $CP$ and the deviation from thermal equilibrium are practically too small.
There exists a large number of mechanisms that fulfil the conditions $i)$-$iii)$ in extensions of the SM, most of which are unfortunately not experimentally testable because they involve supermassive particles and extreme temperatures.
A particularly economic solution to the baryogenesis problem is given by \emph{leptogenesis} \cite{Fukugita:1986hr}. Consider the minimal extension of the SM Lagrangian $\mathcal{L}_{SM}$ by $n$ heavy right handed neutrinos $\nu_{R i}$,
\begin{eqnarray}
\label{eq:Lagrangian}
{\cal L}=\mathcal{L}_{\rm SM}
+\overline{\nu_{{\rm R} i}}{\rm i} \partial\!\!\!/ \nu_{{\rm R} j}
-\frac{1}{2}\overline{\nu_{{\rm R} i}^c}(M_M)_{ij} \nu_{{\rm R} j} 
-\frac{1}{2}\overline{\nu_{{\rm R} i}}(M_M)_{ji}^* \nu_{{\rm R} j}^c
-Y_{ia}^*\overline{\ell_a}\varepsilon\phi \nu_{{\rm R} i}
-Y_{ia}\overline{\nu_{{\rm R} i}}\phi^\dagger \varepsilon^\dagger \ell_a.
\end{eqnarray}
The the superscript $c$ denotes charge conjugation. The $\nu_{R i}$ interact with the SM solely through their Yukawa interactions $Y_{i a}$ to the SM lepton doublets $\ell_a$ ($a=e,\mu,\tau$) and the Higgs field $\phi$, where  $\varepsilon$ is the antisymmetric ${\rm SU}(2)$-invariant tensor with $\varepsilon^{12}=1$.
The model (\ref{eq:Lagrangian}) allows to relate $\eta_B$ to the observed neutrino masses $m_i$ and the parameters in the light neutrino mixing matrix $V_\nu$ \cite{Fukugita:1986hr}: The same particles $\nu_R$ that generate the neutrino masses via the seesaw mechanism \cite{Minkowski:1977sc}
can produce a leptonic matter-antimatter asymmetry via their $CP$ violating interactions in the early universe, which is transferred into a BAU via electroweak sphaleron processes \cite{Kuzmin:1985mm}. 
The light and heavy neutrino mass eigenstates after electroweak symmetry breaking can be expressed in terms of the Majorana spinors
\begin{equation}\label{LightMassEigenstates}
\upnu_i=\left[ V_\nu^{\dagger}\nu_{\rm L}-U_\nu^{\dagger}\theta\nu_{\rm R}^c + V_\nu^{T}\nu_{\rm L}^c-U_\nu^{T}\theta\nu_{\rm R} \right]_i \ , \
N_i=\left[V_N^\dagger\nu_{\rm R}+\Theta^{T}\nu_{\rm L}^{c} +  V_N^T\nu_{\rm R}^c+\Theta^{\dagger}\nu_{\rm L}\right]_i\,.
\end{equation}
with $\theta=m_D M_M^{-1}$ and $m_D=v Y^\dagger$,
where $v$ is the Higgs field expectation value. 
Their interaction strengths are characterised by the quantities
$U_{a i}^2=|\Theta_{ai}|^2$, where $\Theta=\theta U_N^*$.
The unitary matrix $U_N$ diagonalises the heavy neutrino mass matrix 
$M_N=M_M + \frac{1}{2}(\theta^{\dagger} \theta M_M + M_M^T \theta^T \theta^{*})$, 
and $V_N= (1-\frac{1}{2}\theta^T\theta^*)U_N$.
In view of the perspectives to measure the $CP$ violation in neutrino oscillations in experiments, such as NOvA and DUNE, it is of great interest  whether the observable $CP$ violation can be related to the BAU via condition $ii)$.

\section{CP violation and leptogenesis}
From a theoretical viewpoint, the $CP$ violation in a model is best characterised by reparametrisation invariant quantities.
For the model (\ref{eq:Lagrangian}), this has e.g.\ been done explicitly in Ref.~\cite{Hernandez:2015wna}. In the context of the present discussion it is, however, useful to use the Casas-Ibarra parametrisation \cite{Casas:2001sr}
\begin{equation}\label{CasasIbarraDef}
Y^\dagger=\frac{\ii}{v}U_\nu\sqrt{m_\nu^{\rm diag}}\mathcal{R}\sqrt{M_M^{\rm diag}}\,.
\end{equation}
Here $(m_\nu^{\rm diag})_{ij}=\delta_{ij} m_i$ is the light neutrino mass matrix, 
the heavy neutrino masses are given by $(M_M^{\rm diag})_{ij}=\delta_{ij} M_i$ 
and $\mathcal{R}$ is an arbitrary matrix with $\mathcal{R}\mathcal{R}^T=1$.
We in the following work in a flavour basis where $M_M$ is diagonal, i.e., $M_M=M_M^{\rm diag}$.
Note that the number of non-vanishing eigenvalues $m_i$ cannot be larger than $n$, i.e.,
the lightest neutrino is massless if $n=2$ ($m_{\rm lightest}=0$). 
The light neutrino mixing matrix $V_\nu$ can be expressed as $V_\nu= (1-\frac{1}{2}\theta\theta^{\dagger})U_\nu$, where the unitary matrix $U_\nu$ diagonalises the matrix 
\begin{equation}\label{SeesawRelation}
m_\nu = m_D M_M^{-1}m_D^T = v^2 Y^\dagger M_M^{-1} Y^*
\end{equation}
and can be factorised as
\begin{equation}
\label{PMNS}
U_\nu=V^{(23)}U_\delta V^{(13)}U_{-\delta}V^{(12)}{\rm diag}(e^{\ii \alpha_1/2},e^{\ii \alpha_2 /2},1)\,,
\end{equation}
with $U_{\pm \delta}={\rm diag}(1,e^{\mp \ii \delta/2},e^{\pm \ii \delta /2})$.

It is well-known that all dependencies of $\eta_B$ on the phases in $U_\nu$ cancel if the couplings $Y_{ia}$ are of order unity. In this case (\ref{SeesawRelation}) implies $M_i>10^{14}$ GeV, and the BAU is generated in the decay of $\nu_R$ particles at temperatures $T>10^{12}$ GeV, at which the SM flavours are indistinguishable, and $\eta_B$ is independent of the phases in $U_\nu$. For smaller values of $M_i$ and $T<10^{12}$ GeV, the charged lepton Yukawa couplings affect the evolution of leptonic asymmetries. The resulting flavour effects avoid the complete cancellation, and $\eta_B$ depends on the phases in $U_\nu$ \cite{Barbieri:1999ma}.
However, in general there are additional $CP$ violating phases in the matrix $\mathcal{R}$ that affect the BAU, so that $\eta_B$ cannot be uniquely related to the $CP$ violation in $U_\nu$. Since $\mathcal{R}$ is not a fundamental quantity, there is no reason why it should be real or have any other special properties. Moreover, the generation of the observed $\eta_B$ from the decay of right handed neutrinos requires $M_i>10^6$ GeV \cite{Antusch:2009gn}
unless their mass spectrum is highly degenerate \cite{Pilaftsis:2003gt}, which makes a direct detection in experiments impossible in foreseeable time.
These issues are discussed in more detail in the reviews \cite{Buchmuller:2005eh}.

\begin{figure}
	\centering
	\includegraphics[width=0.45\textwidth]{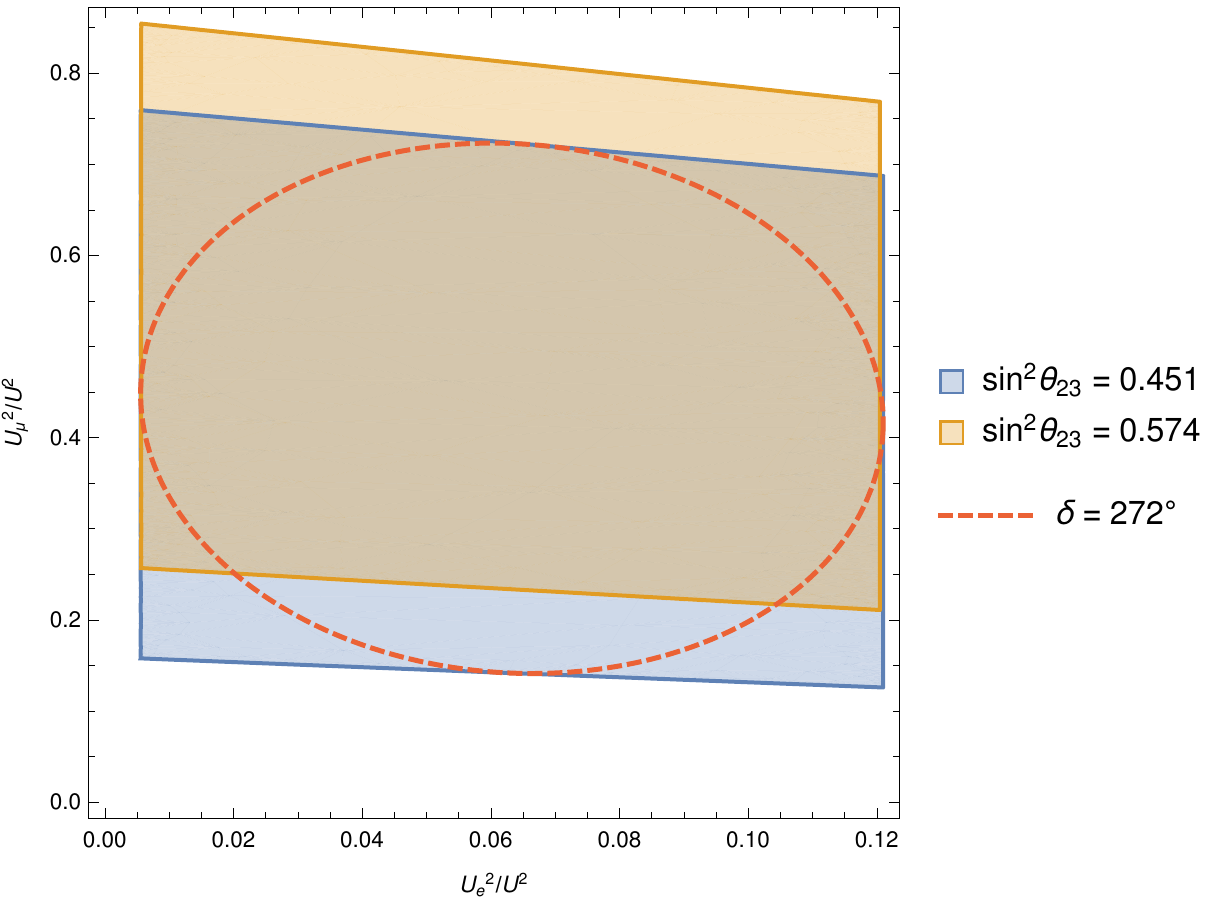}
	\includegraphics[width=0.45\textwidth]{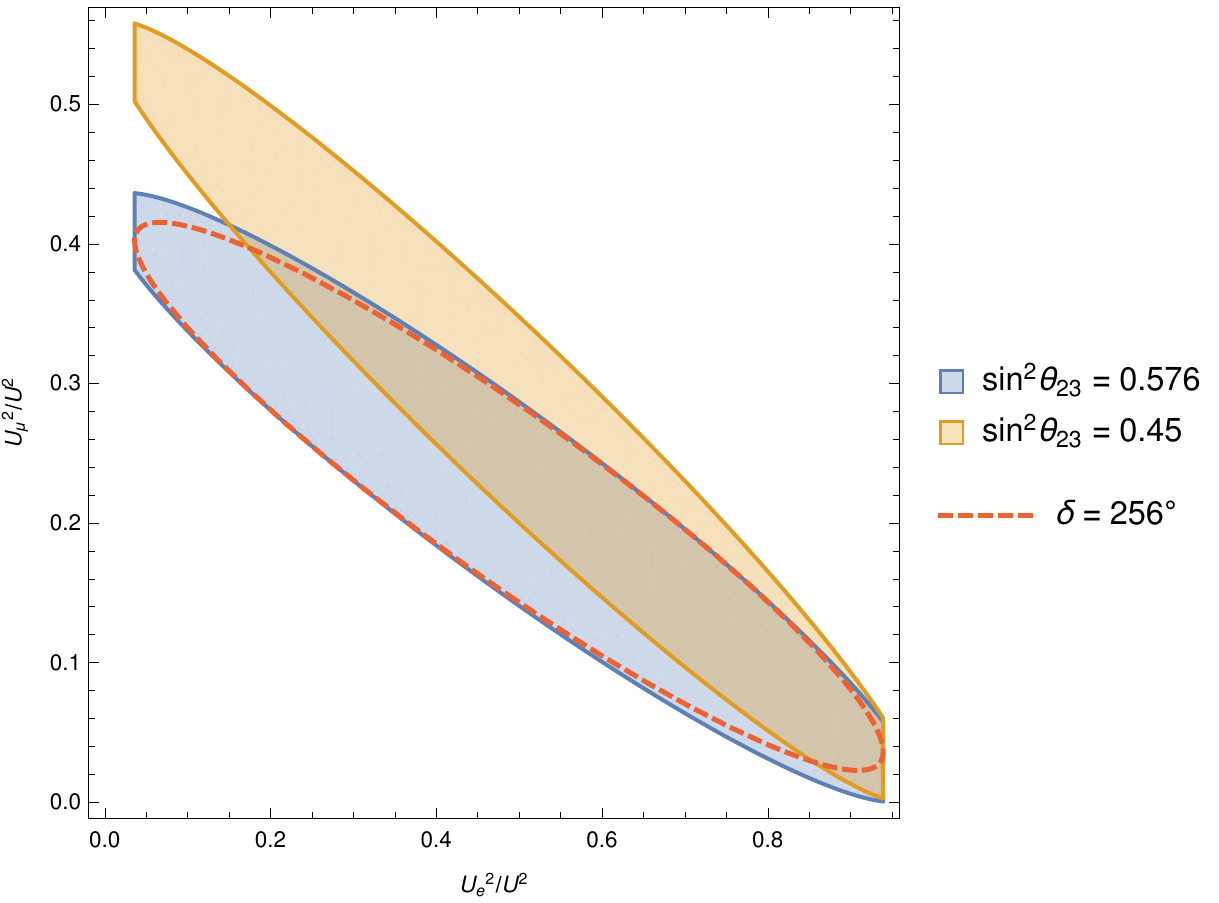}
	\caption{The coloured regions indicate the allowed range of $U_a^2/U^2$ that can be realised by varying the phases in $U_\nu$ after fixing the light neutrino mass splittings and mixing angles to their best fit values for normal hierarchy (left) and inverted hierarchy (right) of light neutrino masses. The difference between the orange and blue region illustrates the change of the predictions if one varies these parameters within their experimental uncertainties. If the Dirac phase $\delta$ is measured independently in light neutrino oscillation experiments, the two dimensional regions will reduce to ``rings'' as illustrated by  the red dashed lines.}
	\label{fig:regions_NO.pdf}
\end{figure}

In the following we focus on an alternative scenario, in which the BAU is not generated in the decay of the right handed neutrinos (\emph{freeze out scenario} or ``thermal leptogenesis''), but in $CP$ violating oscillations during their production (\emph{freeze in scenario} or ``baryogenesis from neutrino oscillations'') \cite{Akhmedov:1998qx,Asaka:2005pn}.
This scenario is feasible for masses $M_i$ below the electroweak scale and is therefore experimentally testable.
The two scenarios simply correspond to different parameter choices in the model (\ref{eq:Lagrangian}) that lead to different realisations of the nonequilibrium condition $iii)$. For superheavy $M_i$, the right handed neutrinos come into equilibrium, freeze out and decay long before electroweak sphalerons freeze out at $T\simeq  130$ GeV. 
In this case the final asymmetry $\eta_B$ that can be observed today is created in the decay of the lightest right handed neutrino.
For $M_i$ below the electroweak scale, the relation (\ref{SeesawRelation}) implies that at least some of the $Y_{ai}$ are much smaller than the electron Yukawa coupling. In this case condition $iii)$ is fulfilled because the heavy neutrinos do not reach thermal equilibrium before sphaleron freezeout. 
In general, the dependence of $\eta_B$ on the various model parameters is rather complicated and can only be calculated numerically because the BAU is generated due to a complex interplay of coherent flavour oscillations and decoherent scatterings. 
Explicit analytic expressions that are valid in certain regimes are e.g.\ given in Refs.~\cite{Shaposhnikov:2008pf,Hernandez:2015wna,Drewes:2016gmt}.
The crucial point in the context of the present discussion is that the BAU in the freeze-in scenario is usually generated at temperatures $T\gg M_i$, when the violation of the total lepton number $L$ by the Majorana masses $M_i$ is suppressed, and the BAU in good approximation originates from a purely flavoured source. Since the couplings of the right handed neutrinos to individual SM flavours $a=e, \mu, \tau$ are determined by the phases in $U_\nu$, cf.\ Fig.~\ref{fig:regions_NO.pdf}, $\eta_B$ is rather sensitive to these phases. In fact, the observed BAU can be generated from the $CP$ violation in $U_\nu$ alone \cite{Canetti:2012kh,Drewes:2015jna}.

\begin{figure}
	\includegraphics[width=0.45\textwidth]{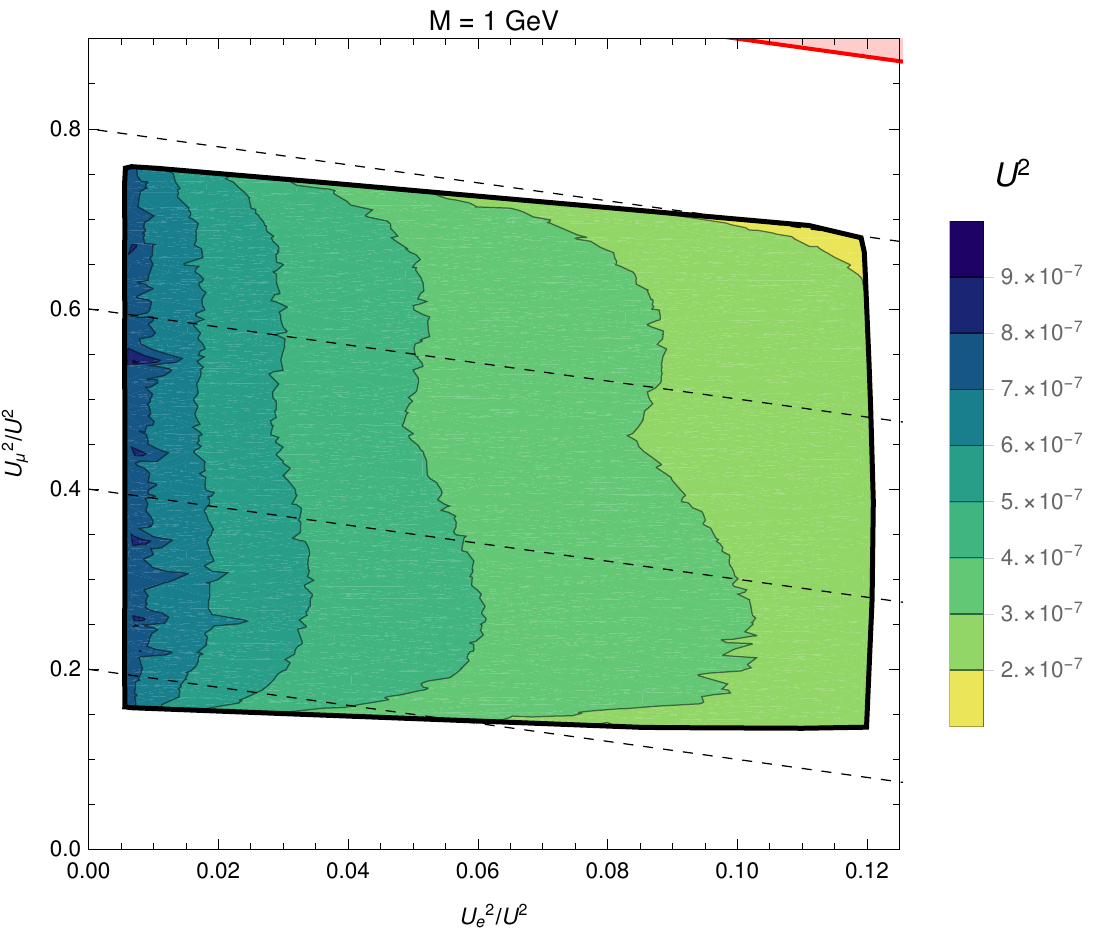}\includegraphics[width=0.45\textwidth]{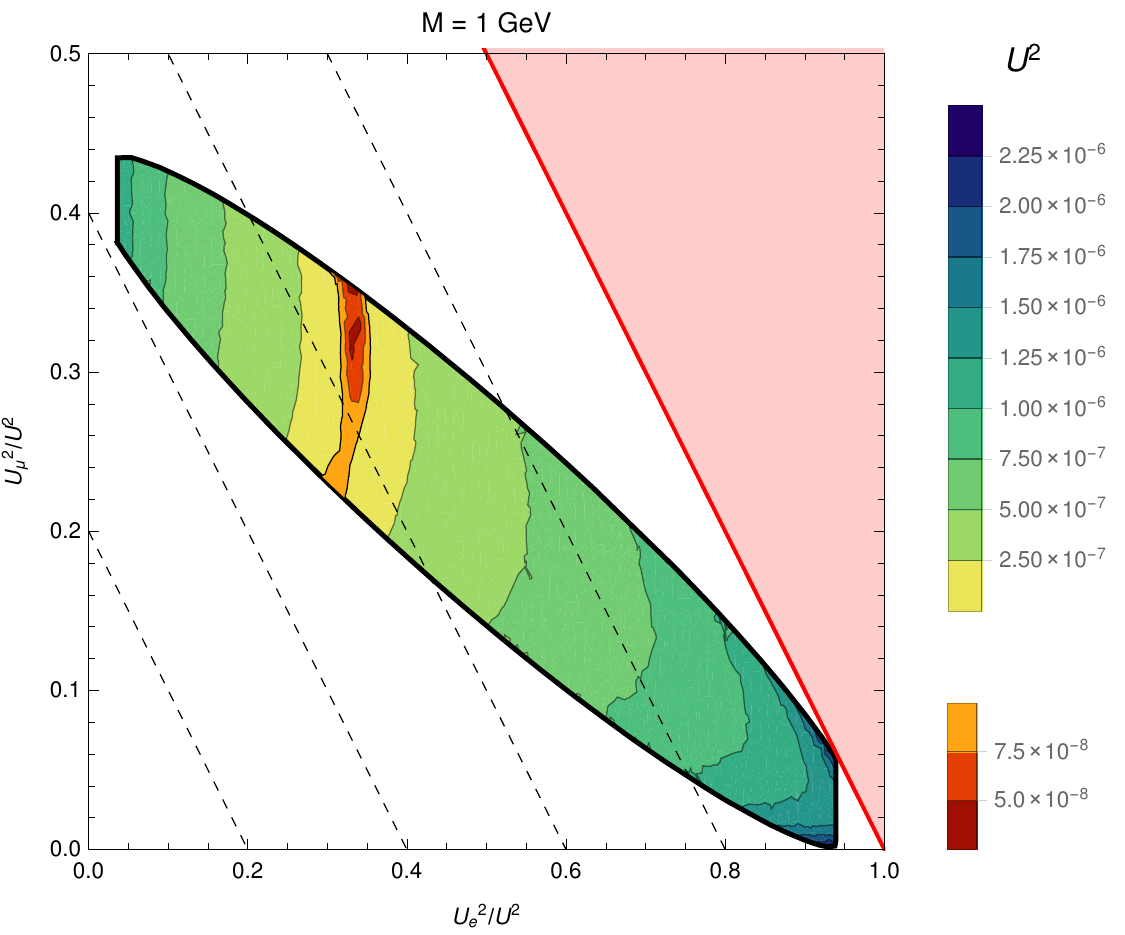}
	\caption{Values of $U_a^2/U^2$ inside the black line are consistent with neutrino oscillation data for normal hierarchy (left) and inverted hierarchy (right) of light neutrino masses, see Fig.~\ref{fig:regions_NO.pdf}. The dashed lines correspond to constant $U_\tau^2$, the light red region is unphysical because it would require $U_\tau^2<0$.  
The coloured regions indicate the maximally allowed $U^2$ for given $U_a^2/U^2$ if one requires that the the observed $\eta_B$ can be generated by leptogenesis with $\bar{M}=1$ GeV. 
	\label{Fig:Lepto_NO}}
\end{figure}

\section{Experimental perspectives}
In the following we consider the minimal seesaw model (\ref{eq:Lagrangian}) with $n=2$ and masses $M_i$ below the electroweak scale. 
Here we closely follow Ref.~\cite{Drewes:2016jae}, but it should be pointed out that various authors have studied the relation between the model parameters, neutrino oscillations and the baryon asymmetry of the universe, see e.g. Refs.~\cite{Shaposhnikov:2008pf,Ruchayskiy:2011aa,Asaka:2011pb,Hernandez:2015wna,Abada:2015rta,Hernandez:2016kel,Drewes:2016lqo,Asaka:2016zib} and references therein.
We use the parametrisation
\begin{eqnarray}
\mathcal{R}^{\rm NH}=
\left(
\begin{tabular}{c c}
$0$ & $0$\\
$\cos \omega$ & $\sin \omega$ \\
$-\xi \sin \omega$ & $\xi \cos \omega$
\end{tabular}\right)
\ , \
\mathcal{R}^{\rm IH}=
\left(
\begin{tabular}{c c}
$\cos \omega$ & $\sin \omega$ \\
$-\xi \sin \omega$ & $\xi \cos \omega$ \\
$0$ & $0$
\end{tabular}\right)
\end{eqnarray}
where $\xi=\pm 1$ and $\omega={\rm Re}\omega+\ii{\rm Im}\omega$ is the complex mixing angle. The index ``NH'' and ``IH'' indicates normal or inverted light neutrino mass hierarchy.
Note that $\alpha_1$ is not physical for $n<3$, so we set $\alpha_1=0$ and $\alpha_2=\alpha$ in the following.
For $n=2$, successful leptogenesis requires that $\Delta M = (M_1-M_2)/2 \ll \bar{M}=(M_1+M_2)/2$ \cite{Asaka:2005pn}. The perspectives for an experimental discovery of the $N_i$ are most promising for large $U_{ai}^2$. Both conditions can be naturally realised in scenarios of an approximate $B-L$ symmetry, which are characterised by small parameters $\mu= \Delta M/\bar{M}$ and $\epsilon=e^{-2\Im \omega}$.
We understand the parameters $\mu$ and $\epsilon$ as phenomenological, specific models can motivate particular trajectories  in the $\mu$-$\epsilon$ plane along which the $B-L$ conserving limit $\mu,\epsilon\rightarrow 0$ should be approached.

\paragraph{Direct searches for $N_i$ -} Heavy neutrinos $N_i$ with masses below the electroweak scale can be 
can be searched for at both, hadron
and lepton colliders.
If the $N_i$ are lighter than the $B$ mesons, they can also be searched for at $B$ factories 
or in fixed target experiments, 
including NA62,
T2K, 
the SHiP experiment proposed at CERN 
or a similar setup proposed at the DUNE beam at FNAL. 
Detailed references on past and planned experimental searches can e.g.\ be found in Refs.~\cite{Atre:2009rg,Ruchayskiy:2011aa,Drewes:2016jae}.
The overall strength $U^2=\sum_a U_a^2$ of the $N_i$ interactions is independent of the choice of $\xi$ and
the phases $\alpha$ and $\delta$ in the matrix $U_\nu$,
\begin{eqnarray}\label{U2NO}
U^2&=&\frac{\Delta M}{M_1 M_2} (m_2-m_3)\cos(2 {\rm Re}\omega)+\frac{\bar{M}}{M_1 M_2}(m_2+m_3)
\frac{1}{2}\left(\epsilon + \frac{1}{\epsilon}\right)
\ {\rm for} \ {\rm NO}\\
U^2&=&\frac{\Delta M}{M_1 M_2} (m_1-m_2)\cos(2 {\rm Re}\omega)+\frac{\bar{M}}{M_1 M_2}(m_1+m_2)
\frac{1}{2}\left(\epsilon + \frac{1}{\epsilon}\right)
 \ {\rm for} \ {\rm IO}\label{U2IO}
.\end{eqnarray}
The phases $\delta$ and $\alpha$ determine the relative sizes of the $N_i$ couplings to individual SM favours.
If the mass resolution of a future experiment that looks for $N_i$ is smaller than $\Delta M$, then the mixings $U_{a1}^2$ and $U_{a2}^2$ can be measured independently. 
This would e.g.\ be possible at the SHiP experiment or a future lepton collider.  
The $U_{ai}^2$ are invariant under changes $(\delta,\alpha,{\rm Re}\omega)\rightarrow (-\delta,2\pi-\alpha,-{\rm Re}\omega)$.\footnote{This is in addition to the (unphysical) parameter degeneracy that comes from the inherent ``multiple coverage'' of the parameter space in the Casas Ibarra parametrisation: 
${\rm Re}\omega\rightarrow {\rm Re}\omega+\pi$ changes the sign of all $Y_{ia}$, which can always be compensated by  field redefinitions $N_i\rightarrow -N_i$.
Adding $2\pi$ to $\alpha$ while swapping the sign of $\xi$ leaves the $Y_{ia}$ invariant for inverted hierarchy and swaps the sign for normal hierarchy.
Swapping the signs of $\xi$, ${\rm Im}\omega$, $\Delta M$ and changing ${\rm Re}\omega\rightarrow \pi-{\rm Re}\omega$ swaps the labels of $N_1$ and $N_2$, with no physical consequences.
}
Hence, a measurement of all $U_{ai}^2$ would fix the phases $(\delta,\alpha)$ and $\omega$ up to one discrete transformation, with $\bar{M}$ and $\Delta M$ being extracted from the kinematics.
The remaining parameter degeneracy can be broken by an independent measurement of $\delta$, see below.\footnote{In the first arxiv preprints of this contribution and Ref.~\cite{Drewes:2016jae} we missed the invariance of the $U_{ai}^2$ under the transformation $(\delta,\alpha,{\rm Re}\omega)\rightarrow (-\delta,2\pi-\alpha,-{\rm Re}\omega)$ and incorrectly claimed that all model parameters can be reconstructed from the $U_{ai}^2$ alone.}

If the experimental resolution is worse than $\Delta M$, then the mixings $U_{a1}^2$ and $U_{a2}^2$ cannot be measured independently. Instead, experiments are sensitive to $U_a^2=\sum_i U_{ai}^2$, which are invariant under one more transformation that has no simple analytic form.\footnote{Details will be provided in an updated version of Ref.~\cite{Drewes:2016jae}.}
Moreover, for $\mu\rightarrow0$, the $U_a^2$ become independent of the unknown parameter ${\rm Re}\omega$. 
As a result, one cannot put meaningful constraints on the parameters $\Delta M$ and ${\rm Re}\omega$, which are crucial for leptogenesis, if $\mu$ is very small. 
However,  consistency checks for both, the hypothesis that the $N_i$ generate the light neutrino masses via the seesaw mechanism and the BAU via leptogenesis, are still possible for arbitrarily small $\mu$.
Since $U^2$ is independent of $\xi$, a measurement of all $U_a^2$ allows to uniquely fix ${\rm Im}\omega$ for $\mu=0$.
If in addition $\epsilon\ll1$, then the ratios $U_a^2/U^2$ are in good approximation independent of $\omega$ and $\bar{M}$, and are entirely determined by the phases $\alpha$ and $\delta$ alone. This is illustrated in Fig.~\ref{fig:regions_NO.pdf}.
Since not every set $(U_e^2,U_\mu^2,U_\tau^2)$ can be realised by varying $(\alpha,\delta,{\rm Im}\omega)$, see Fig.~\ref{fig:regions_NO.pdf}, 
the consistency of the $U_a^2$ measurements with each other and with a possible determination of $\delta$ in neutrino oscillation experiments
provide a powerful test of whether the discovered particles are really part of minimal seesaw model of neutrino mass generation.

The requirement to reproduce the observed $\eta_B$ (in addition to the observed neutrino masses) further restricts the range of allowed $(\delta,\alpha)$ for given $U^2$. This is only partly related to the role of $(\delta,\alpha)$ as sources of $CP$ violation, but predominantly owed to the fact that these phases fix the ratios $U_a^2/U^2$: For values of $U^2$ that are large enough to be experimentally accessible, 
the washout of lepton asymmetries in the early universe tends to erase the BAU before sphaleron freezeout unless there is a hierarchy amongst the $U_a^2$ that leads to a flavour asymmetric washout and allows the asymmetry in one flavour to survive. This restricts the range of $U_a^2/U^2$ that are consistent with leptogenesis for given $U^2$, see Fig.~\ref{Fig:Lepto_NO}.

\paragraph{Measurements of the Dirac phase -}
A measurement of a nonzero $\delta$ would not only be a clear proof that $CP$ violation exists in the lepton sector, which is a key requirement for leptogenesis, but also allow to make predictions about the heavy neutrino flavour mixing pattern $U_a^2/U^2$, see Fig.~\ref{fig:regions_NO.pdf}. 
If heavy neutrinos are found in collider or fixed target experiments, then the consistency of the observed value of $\delta$ with the indirect constraints extracted from the active-sterile mixings $U_a^2$ provides a very strong test of the hypothesis that these particles give masses to the light neutrinos.
If $\Delta M$ is large enough to be resolved kinematically and all $U_{ai}^2$ can be measured individually, then an independent measurement of $\delta$ allows to uniquely fix all model parameters and reconstruct the Lagrangian (\ref{eq:Lagrangian}).    
This would allow to predict the rate of neutrinoless double $\beta$ decay, the outcome of future searches for violation of lepton number or lepton universality and the amount of unitarity violation in $V_\nu$. Moreover, it would also allow to predict the value of $\eta_B$. Hence, the hypotheses that the $N_i$ are the origin of neutrino masses and matter in the universe are both fully testable.

\paragraph{$CP$ violation in $N_i$ decays -} 
In Ref.~\cite{Cvetic:2014nla},
it has been pointed out that the $CP$ violation in $N_i$-mediated
decays of charged pseudoscalar mesons $M$ 
into lighter mesons $M'$ and leptons could be measured, which would give direct access to the $CP$ violation in $N_i$ decays (assuming that the initial ``production asymmetry'' between positively and negatively charged $M$ is under control).
This $CP$ asymmetry $\mathcal{A}_{CP}\equiv \frac{\sum_i \Gamma(M^-\to \ell_a^-\ell_b^- M'^+)-\Gamma(M^+\to \ell_a^+\ell_b^+ M'^-)}{\sum_i \Gamma(M^-\to \ell_a^-\ell_b^- M'^+)+\Gamma(M^+\to \ell_a^+\ell_b^+ M'^-)}$ can be approximated as
$\mathcal{A}_{CP}\approx\frac{y}{y^2+1}\sin\left({\rm arg}\left[\frac{\theta_{a2}}{\theta_{a1}}\frac{\theta_{b2}}{\theta_{b1}}\right]\right)$,
where $y=\Delta M /\bar{\Gamma}_N$ and $\bar{\Gamma}_{N} = \frac{G_F^2 \bar{M}^5}{96 \pi^3}\mathcal{N} U^2$ with $\mathcal{O}(\mathcal{N})=10$ is a measure of the $N_i$ decay width.
For $y\sim 1$ the $CP$ violation can become large enough to be measurable. 
The number of expected events also depends on some effective branching ratio 
${\rm Br}_{\rm eff}\sim 10^2 U^4$.
Observing the $CP$ violation requires $|\mathcal{A}_{CP}|{\rm Br}_{\rm eff}$ to exceed the inverse of the number of produced mesons in the experiment. In the leptogenesis parameter region (where $\mu\ll 1$) this is experimentally difficult: For given $\bar{M}$, ${\rm Br}_{\rm eff}$ can only be increased by decreasing $\epsilon$, but the phase difference in $\mathcal{A}_{CP}$  approaches $\pi/2$ in the limit $\mu,\epsilon\rightarrow 0$.
Hence, in the minimal model (\ref{eq:Lagrangian}) with $n=2$, it would be extremely challenging to measure the $CP$ violation in the parameter region where it can be responsible for the BAU. However, the perspectives may be better for scenarios with $n>2$, which contain more unconstrained $CP$ violating phases \cite{Zamora-Saa:2016qlk}. 

\paragraph{Neutrinoless double $\beta$ decay -} 
For 
$n=2$ and  $\mu,\epsilon\ll 1$,
the effective Majorana mass $m_{\beta\beta}$ that governs the rate of the $0\nu\beta\beta$ decay can be expressed as
\begin{eqnarray}
m_{\beta\beta}^{\rm NH}&\simeq& \left| [1-f_A(\bar{M})]m_{\beta\beta}^\nu+f_A^2(\bar{M})\frac{\bar{M}^2}{\Lambda^2}
\frac{\mu}{\epsilon}
|m_{\rm atm}|
e^{-2\ii(\Re\omega+\delta)}\right|,\nonumber \\ \nonumber
m_{\beta\beta}^{\rm IH}&\simeq& \bigg| [1-f_A(\bar{M})]m_{\beta\beta}^\nu+f_A^2(\bar{M})\frac{\bar{M}^2}{\Lambda^2}
\frac{\mu}{\epsilon}
|m_{\rm atm}| \cos^2\uptheta_{13}
e^{-2\ii \Re\omega}(\xi e^{\ii \alpha_2/2}\sin \uptheta_{12}+\ii e^{\ii\alpha_1/2}\cos \uptheta_{12})^2
\bigg|,\label{oderplus}
\end{eqnarray}
where $f_A(M)\simeq \frac{\Lambda^2}{\Lambda^2+\bar{M}^2}$, $m_{\beta\beta}^\nu=\sum_i (U_\nu)_{ei}^2m_i$ is the contribution from light neutrino exchange  and  $\Lambda^2$ is the momentum exchange in the decay. 
While it is well-known that the contribution from $N_i$ exchange to the $0\nu\beta\beta$ decay can be sizeable \cite{Bezrukov:2005mx,Blennow:2010th,Asaka:2011pb},
it was long believed that the requirement $\mu\ll 1$ in the context of leptogenesis implies that $m_{\beta\beta}\simeq | [1-f_A(\bar{M})]m_{\beta\beta}^\nu|$, and $m_{\beta\beta}$ is insensitive to the heavy neutrino parameters (except $\bar{M}$) \cite{Bezrukov:2005mx,Asaka:2011pb}.
Recently it has been pointed out that there exists a corner in the parameter space of the seesaw model with $n=2$ in which the observed $\eta_B$ can be reproduced while the term $\propto \mu/\epsilon$ in $m_{\beta\beta}$ dominates \cite{Drewes:2016lqo,Hernandez:2016kel,Asaka:2016zib}. This implies that the parameter ${\rm Re}\omega$ can be constrained from $0\nu\beta\beta$ decay even if $\mu$ is so small that direct searches cannot kinematically distinguish the two heavy neutrinos.
This of course requires that $\Lambda/\bar{M}$ is not too small. For $n=3$, there is no requirement for a mass degeneracy  for successful leptogenesis \cite{Drewes:2012ma}, and a large $m_{\beta\beta}$ can easily be made consistent with the observed $\eta_B$ \cite{Drewes:2016lqo}.

\section{Conclusions}
In the type I seesaw model with a low seesaw scale, the $CP$ violation in the light neutrino mixing matrix is closely related to the properties of the heavy neutrinos and the baryon asymmetry of the universe. A measurement of the Dirac phase $\delta$ would allow to make testable predictions for the couplings of the heavy neutrinos to individual SM flavours. 
If any heavy neutral leptons are experimentally discovered in the future, all model parameters can be reconstructed from measurements of $\delta$ and the mixings $U_{ai}^2$, making the low scale seesaw a fully testable mechanism of neutrino mass generation and baryogenesis.

\section*{Acknowledgements}
This research  was  supported  by  the  DFG  cluster  of  excellence  'Origin  and  Structure  of  the Universe' (www.universe-cluster.de).


\end{document}